\documentclass[pra,twocolumn,superscriptaddress]{revtex4-1}
\usepackage{float,graphicx,amsmath,dsfont,verbatim,amsthm}
\usepackage{physics}
\usepackage{xcolor,braket}

\newtheorem{definition}{Definition}

\newcommand{\etal}{\textit{et al}.~}

\usepackage{xcolor}
\definecolor{maroon}{RGB}{100,20,20}
\definecolor{dblue}{RGB}{20,20,100}
\usepackage[colorlinks=true,linkcolor=dblue,
citecolor=maroon,urlcolor=maroon]{hyperref}
\begin{document}
\title{No contextual advantage in non-paradoxical scenarios of two state vector formalism}
\author{Jaskaran Singh}
\email{jsinghiiser@gmail.com}
\affiliation{Departamento de F\'{\i}sica Aplicada II, Universidad de Sevilla, E-41012 Sevilla, Spain}
\author{Rajendra Singh Bhati}
\email{ph16076@iisermohali.ac.in}
\affiliation{Department of Physical Sciences, Indian
Institute of Science Education and Research (IISER)
Mohali, Sector 81 SAS Nagar, Manauli PO 140306 Punjab India.}
\author{Arvind}
\email{arvind@iisermohali.ac.in}
\affiliation{Department of Physical Sciences, Indian
Institute of Science Education and Research (IISER)
Mohali, Sector 81 SAS Nagar, Manauli PO 140306 Punjab India.}
\affiliation{Vice Chancellor, Punjabi University Patiala,
147002, Punjab, India}
\begin{abstract}
The two state vector formalism (TSVF) was proposed by Aharonov,
Bergmann, and Lebowitz (ABL) to provide a way for the
counterfactual assignment of the probabilities of outcomes
of contemplated but unperformed measurements on quantum
systems. This formalism underlies various aspects of foundations of quantum theory
and has been used significantly in the development of weak values and several
proofs of quantum contextuality.  
We consider the
application of TSVF, with pre- and post-selection (PPS) and the
corresponding ABL rule, as a means to unearth quantum contextuality. 
We use the principle of exclusivity to classify the resultant pre- and post-selection 
scenarios as either paradoxical or non-paradoxical. In light of this, we find that 
several previous proofs of the emergence of contextuality in PPS scenarios are 
only possible if the principle of exclusivity is violated and are therefore classified as 
paradoxical. We argue that these do not constitute a proper test of contextuality.
Furthermore, we provide a numerical analysis for the KCBS scenario as applied in the 
paradigm of TSVF and find that non-paradoxical scenarios do not offer any contextual 
advantage. Our approach can be easily generalized for other contextual scenarios as well.
\end{abstract}
\maketitle
\section{Introduction}
The standard quantum theory does not provide a framework for
making predictions about the measurements in the past
(\textit{retrodiction}) of a quantum system once it has been
measured in a definite state.  Aharonov, Bergmann, and
Lebowitz (ABL), in their seminal work on the time symmetry
in successive quantum measurements, introduced a
reformulation of the standard quantum theory where one can
meaningfully talk about statistical predictions of a
measurement on a pre-and post-selected (PPS) ensemble at an
intermediate time \cite{PhysRev.134.B1410}.  The
\textit{retrodiction} formula derived by ABL (ABL rule) is
the probability of a measurement outcome conditioned on
the outcomes of a preceding and a succeeding
measurement.

A generalized framework for PPS ensembles in terms of `weak
values' was formulated as two state vector formalism
(TSVF)~\cite{PhysRevLett.60.1351,PhysRevA.41.11,Aharonov-1991}
in order to experimentally validate the ABL
formulation~\cite{PhysRevLett.66.1107,PhysRevLett.94.220405}.
In TSVF, the complete description of a quantum system is
specified by two state vectors, one evolving forward in time
and the other one evolving backwards. Here the arrow of time is
described by the order of preceding and succeeding measurements.
TSVF finds intriguing applications in quantum
foundations~\cite{PhysRevLett.74.2405,Kocsis1170,Mahlere1501466,
aharanov_hardy,PhysRevLett.102.020404,PhysRevLett.100.026804,Lundeen2011}
and quantum information processing~\cite{RevModPhys.86.307}.
The ABL retrodiction, more specifically the TSVF, has
resulted in various counter intuitive results commonly
called PPS paradoxes~\cite{RESCH2004125,PhysRevA.87.052104,PhysRevLett.111.240402,
Aharonov2017,1367-2630-15-11-113015,Denkmayr2014,Bancal2013,
Das_2020,Liu2020}.  In a recent work~\cite{BHATI2022127955}
counterfactual use of ABL retrodiction has been shown to run
into a direct contradiction with operational quantum
mechanics challenging the completeness of TSVF.  Therefore
further investigations on the appropriateness of ABL
retrodiction in connection with various non-classical
aspects of quantum theory is critical in order to
pinpoint the exact role of TSVF in the studies of quantum
foundations.

Contrary to the Born rule, probabilities assigned by
the ABL
formula are determined by
the specification of
the measurement setting of an observable and
on pre and post selected states in context of which the observable is being
measured.
This sort of context dependency of measurements
has led to connections between  PPS paradoxes and
contextuality~\cite{PhysRevLett.54.5}: since the
probabilities assigned to measurement outcomes are
explicitly context dependent, there is no motivation to
consider a non-contextual hidden variable theory as a
realistic extension of operational quantum theory.
Nevertheless, this reasoning has been disputed based on the
fact that Bell-type correlations can be simulated using
post-selections in local hidden variable
theories~\cite{PhysRevLett.56.2337}.  Therefore, the mere
presence of context dependent elements in ABL formula should
not be sufficient to prove Bell-Kochen-Specker (BKS)
theorem~\cite{Koc,RevModPhys.65.803} or the various statistical versions of contextuality~\cite{PhysRevLett.112.040401}.  It is required to
dive deeper in order to establish a valid connection between
contextuality and the ABL retrodiction formula.

Mermin~\cite{PhysRevLett.74.831} showed the existence of a
strong connection between the two by illustrating how
measurements used in a proof of BKS theorem can give birth
to unsolvable PPS paradoxes indicating a kind of
impossibility of non-contextual hidden variable theories.
Leifer and Spekkens~\cite{PhysRevLett.95.200405} later
logically showed that for every PPS paradox with a
scenario involving non-orthogonal pre-and post-selection
states, there exists an associated proof of BKS theorem.
Their proof is based on the fact that ABL probability
assignments of certain sets of projectors in a variant of
3-box paradox violate algebraic constraints dictated by
classical probability theory.  An exhaustive discussion on
the same in relation with weak values was presented by
Tollaksen~\cite{Tollaksen_2007}. Another important
contribution in this direction was recently made by
establishing a direct connection between anomalous weak
values and contextuality where it has been suggested that
anomalous weak values can be taken as proofs of
contextuality~\cite{PhysRevLett.113.200401,
PhysRevA.100.042116}.
So far the studies in this avenue of research have been
focused on logical proofs of contextuality
invoking only the paradoxical nature of ABL probability
assignments.  In these logical proofs of contextuality one arrives at a contradiction
while making assignment of probabilities to various outcomes following the ABL rule. Such proofs generally
involve the paradoxes generated by the application of the ABL rule.
Therefore, it is natural to ask
whether there is any contextual advantage in
situations with ``non-paradoxical assignments" of ABL
probabilities.

In this paper, by analyzing the Klyachko-Can-Binicio\ifmmode
\breve{g}\else \u{g}\fi{}lu-Shumovsky (KCBS)
scenario~\cite{PhysRevLett.101.020403} (which comprises a statistical
proof of contextuality) with the ABL formula, we show
that non-paradoxical ABL probability
assignments do not give rise to any contextual advantage. Furthermore, in
order to produce an advantage one needs to
renounce the exclusivity principle which is central to
any operational theory~\cite{PhysRevLett.110.060402, PhysRevA.89.030101,PhysRevResearch.2.042001}.  Our result raises serious
questions about the ABL-contextuality connections that have
been advocated by previous authors: is this connection
merely an illusion created by post-selection just like the
detection efficiency loophole in Bell non-locality tests?
Since the paradoxical sector of ABL probabilities requires abandoning the notion of 
principle of exclusivity can
ABL retrodiction be considered an appropriate description of
quantum systems at all?

The paper is organized as follows: In Sec.~\ref{sec:back}
we introduce the concept of PPS scenarios and briefly
discuss the ABL formula and consequently the TSVF. In
Sec.~\ref{sec:result} we present our main result that the
ABL rule is unable to correctly predict the statistics of
the KCBS scenario. In Sec.~\ref{sec:conc} we offer some
concluding remarks.

\section{ABL rule, TSVF and PPS paradoxes}
In this section we describe the TSVF and ABL retrodiction rule. We define PPS scenarios 
and introduce the notion of paradoxical and non-paradoxical
nature of them. This classification depends on whether the probability assignments 
are properly conditioned by the exclusivity principle or not.

\subsection{TSVF and ABL retrodiction}
\label{sec:back}
In this subsection we illustrate a general pre- and
post-selected scenario and elucidate how the ABL rule can be
used to assign probabilities to intermediate measurements.

\begin{figure}
\includegraphics[scale=1]{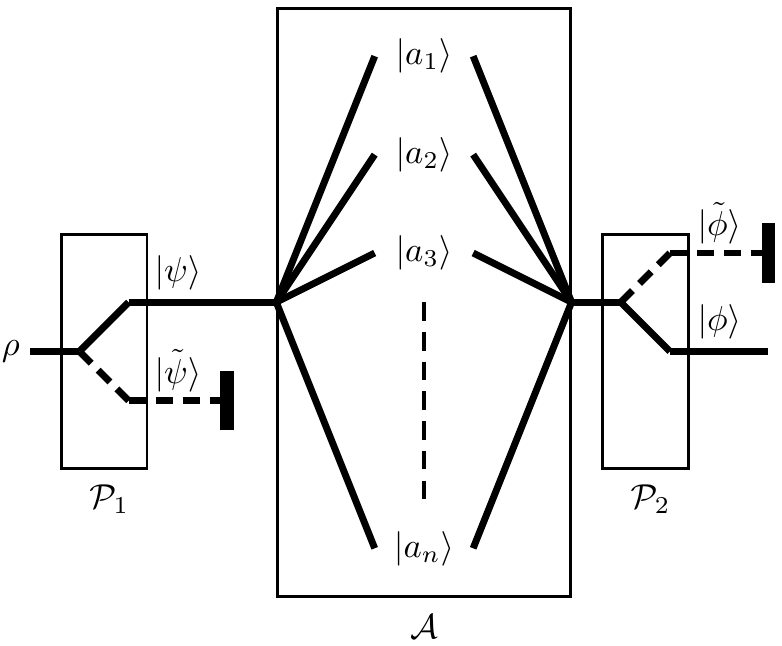}
\caption{A PPS scenario with a
measurement of an observable $\mathcal{A}$ at an intermediate instance of time. System is
pre-selected in state $\ket{\psi}$ by performing a projective
measurement $\mathcal{P}_1$ and filtering the outcome corresponding to the state 
$\ket{\psi}$ and post selected in state
$\ket{\phi}$ by filtering outcome $\tilde{\ket{\phi}}$ of
measurement $\mathcal{P}_2$. $\lbrace\ket{a}_i\rbrace$ is
the set of all possible outcomes of an intermediate
observable $\mathcal{A}$}
\label{fig:PPS}
\end{figure}

A pre- and post-selection scenario deals with statistical
assignment of probabilities to the outcomes of the
measurement of an observable $A$ at time $t$ when the system
is pre-selected to be in the state $\ket{\psi}$ at some time
$t_i<t$ and post-selected in the state $\ket{\phi}$ at a later moment in
time $t_f>t$. Pre-selection is
achieved by performing a projective measurement
$\mathcal{P}_1 \equiv \lbrace
\ketbra{\psi},\mathds{1}-\ketbra{\psi}\rbrace$ at time $t_i$
on the initial state of the system $\rho $ (which can be
chosen arbitrarily) and selecting only the outcomes
corresponding to $\ketbra{\psi}$. Similarly for
post-selection one can perform a projective measurement of
$\mathcal{P}_2 \equiv \lbrace
\ketbra{\phi},\mathds{1}-\ketbra{\phi}\rbrace$ at time $t_f$
where outcomes corresponding to $\ketbra{\phi}$ are filtered
(see Fig.~\ref{fig:PPS}). It is apparent that such
probability assignments are conditioned on PPS states
$\ket{\psi}$ and $\ket{\phi}$ and therefore time symmetric.

Consider an observable $\mathcal{A}$ with outcomes $\lbrace\ketbra{a_i}\rbrace$ which is measured 
after pre-selecting a state $\ket{\psi}$ and afterwards post-selecting a state $\ket{\phi}$. The
probability of obtaining the outcome $\ketbra{a_i}$ conditioned on pre-and post-selections
is given as
\begin{equation}
\zeta_i =
\frac{|\braket{\phi|a_i}|^2|\braket{a_i|\psi}|^2}
{\sum_j|\braket{\phi|a_j}|^2|\braket{a_j|\psi}|^2},
\end{equation}

which can be simplified as

\begin{equation}
\zeta_i =
\frac{|\langle\phi|\Pi_i|\psi\rangle|^2}{\sum_j|\langle\phi|\Pi_j|\psi\rangle|^2},
\label{eq:ablrule1}
\end{equation}

where $\Pi_i=\ketbra{a_i}$. As one can see, $\zeta_i$ for a projector $\Pi_i$ is dependent on 
PPS states $\ket{\psi}$ and $\ket{\phi}$. A different choice of these will yield different 
probability assignments. Furthermore, the measurement context of the projector $\Pi_i$ also plays
a major role. If the set of measurement settings in which $\Pi_i$ appears is chosen differently, the term
$\sum_j|\langle\phi|\Pi_j|\psi\rangle|^2$ will also change. All of the aforementioned choices 
form a context for the projector $\Pi_i$.

This makes ABL formula
given in Eq.~\eqref{eq:ablrule1} inherently context dependent and 
led Albert \etal to draw a parallel between ABl retrodiction
and quantum contextuality.

There are two ways to interpret Eq.~\eqref{eq:ablrule1}.
In the first case, the observable $\mathcal{A}$ is
actually measured after performing a pre-selection.  Post
selection is then performed on the state after the
measurement of $\mathcal{A}$. In this case there are a
total of three different sequential measurements being
performed. This case is known as non-counterfactual
assignment of probabilities
\cite{Kastner1999,MILLER199631,PhysRevA.51.4373}.

In the second case, the observable $\mathcal{A}$ is not actually
measured, but rather a probability distribution over its
outcomes is assigned counterfactually depending on the PPS
states~\cite{Aharonov-1991}.  This case is known as
counterfactual assignment of probabilities. It has been argued by Aharonov and
collaborators that simultaneous counterfactual probability 
assignments, in accordance with Eq.~\eqref{eq:ablrule1},
to any number of arbitrary observables is possible.
This in general produces the various PPS paradoxes, as we will define
in next subsection.

For the remainder of this paper we consider only
counterfactual measurement setting of the observable $\mathcal{A}$.
This rule (Eq.~\eqref{eq:ablrule1}) is eponymously known as
the ABL rule and is the same for both the aforementioned
cases. However, the interpretation for both the cases is
entirely different and leads to some interesting results,
especially when linked to counterfactual assignments of
projectors in the KCBS scenario.

\subsection{Paradoxical and non-paradoxical PPS scenarios}
\label{sec:result}

In this subsection we provide a classification of PPS scenarios into paradoxical and 
non-paradoxical.
We then proceed to show how the
KCBS scenario can be modified to fit within the paradigm of
TSVF and whether the latter can help in predicting the
correct statistics of the former.  We then provide a general
algorithm to check for contextual advantages for other
contextuality scenarios. 
\begin{definition}[Counterfactual PPS scenario]
A counterfactual PPS scenario is specified by $(\langle\phi
|| \psi\rangle,\mathcal{M})$ where $\langle\phi || \psi\rangle$ is a
two-state describing the PPS
ensemble and a projective valued measure (PVM),
$\mathcal{M} = \{\Pi_i\}\,\,( i=1,2\cdots n)$ is the
counterfactual measurement setting at an intermediate time.
The corresponding $\lbrace\zeta_i\rbrace$ given by
Eq.~\eqref{eq:ablrule1} are then the counterfactual
probability assignments for the PVM $\mathcal{M}$.
\end{definition}

For certain pre- and post-selections
counterfactual probability assignments can lead to
paradoxical situations. The 3-box paradox is a
case in point~:
consider a particle that is pre-selected in state $(\ket{A}+
\ket{B}+\ket{C})/\sqrt{3}$ and post-selected in the state
$(\ket{A}+\ket{B}-\ket{C})/\sqrt{3}$, where $\ket{A},
\ket{B}, \ket{C}$ represent the states of the particle
being in boxes $A, B$ and $C$ respectively. Now consider two
possible counterfactual measurement settings
$\mathcal{A}=\{\ketbra{A},\mathds{1}-\ketbra{A}\}$ and
$\mathcal{B}=\{\ketbra{B},\mathds{1}-\ketbra{B}\}$. It is easy to
visualize $\mathcal{A}$ and $\mathcal{B}$ as being the actions of opening
the boxes $A$ and $B$ respectively at an intermediate time
in order to check whether the particle was present there. A
counterfactual probability assignment to both the projectors
$\ketbra{A}$ and $\ketbra{B}$ can be made according to ABL
formula \eqref{eq:ablrule1}. However, it can be seen that such an
assignment leads to a
situation in which the particle is present in box $A$ with unit probability and 
box $B$ with unit probability~\cite{Kastner1999}. 

The exclusivity principle states that 
the sum of probabilities of mutually exclusive events cannot be greater than $1$. Therefore, such scenarios in which two 
mutually exclusive events are assigned unit probabilities is paradoxical. This motivates
the following definition.

\begin{definition}[Logical PPS paradox]
A logical PPS paradox consists of at least two
counterfactual PPS scenarios $(\langle\phi ||
\psi\rangle,\mathcal{M}_1)$ and $(\langle\phi || \psi\rangle,\mathcal{M}_2)$
where $\mathcal{M}_i=\{\Pi_i,\mathds{1} -\Pi_i\}$ for $i=1,2$ and
$\tr(\Pi_1\Pi_2)=0$ such that $\zeta_1+\zeta_2>1$ where
$\zeta_i$ is counterfactual probability assigned to $\Pi_i$.
\end{definition}

The logical PPS paradox is defined for PPS scenarios which violate the 
As shown in Ref.~\cite{PhysRevLett.95.200405}, to every
corresponding logical PPS paradox there exists a proof of
BKS theorem. However, the relation between non-paradoxical
PPS scenarios and contextuality is still left unexplored. To
that end we make the following definition to distinguish
between the paradoxical and non-paradoxical sector of PPS
scenarios.
\begin{definition}[Paradoxical and non-paradoxical sector]
The set of all two-states that generate logical PPS
paradoxes for given two counterfactual measurement settings
$\mathcal{M}_1=\{\Pi_1,\mathds{1}-\Pi_1\}$ and $\mathcal{M}_2=\{\Pi_2,\mathds{1}
-\Pi_2\}$ such that $\tr(\Pi_1\Pi_2)=0$ is called the
paradoxical sector corresponding to pair $\{\mathcal{M}_1,\mathcal{M}_2\}$.
We term the set of all two-states that are not elements of
the above as the non-paradoxical sector corresponding to the
pair $\{\mathcal{M}_1,\mathcal{M}_2\}$. 
\end{definition}

\section{KCBS construction and non-contextuality}

We now analyze whether the non-paradoxical sector of PPS
scenarios can offer a proof of contextuality. We first focus on the minimal proof
of state dependent contextuality, namely the KCBS scenario
and construct an ontological description of the same via the
TSVF.

Consider a scenario consisting of $5$ tests $e_i$,
$i\in\lbrace0,1,2,3,4\rbrace$. A test is an experiment which
yields some statistics for a given preparation. These tests
are assumed to be cyclically exclusive, i.e., 
\begin{equation}
P(e_i)+P(e_{i\oplus 1})\leq 1,
\label{eq:exclusivity}
\end{equation} 
where $i\oplus 1$ is taken modulo $5$. This scenario is
eponymously named as KCBS, in honor of the people who first
studied it. The KCBS scenario can be represented on a graph,
whose vertices correspond to tests and two vertices are
connected by an edge if they are exclusive. This scenario is
capable of revealing quantum contextuality if the following
inequality is violated,
\begin{equation}
\mathcal{K}:=\sum_{i=0}P(e_i)\leq 2,
\label{eq:kcbs}
\end{equation}
where the underlying ontic probability distribution,
$P(e_i)$ is assumed to be non-contextual. This is the well
known KCBS inequality.

A valid construction of the KCBS scenario for the quantum
case is as follows. Consider $5$ different PVMs: $\mathcal{M}_i =
\lbrace\Pi_i,\mathds{1}-\Pi_i\rbrace$,
$i\in\lbrace0,1,2,3,4\rbrace$ and tr$(\Pi_i\Pi_{i\oplus
1})=0$. Each projector corresponds to a test in the KCBS
scenario and cyclic orthogonality ensures the required
exclusivity conditions given in Eq.~(\ref{eq:exclusivity}).
The maximum quantum value of the KCBS inequality~\eqref{eq:kcbs} for
the aforementioned settings and a state $|\psi\rangle$ is
\begin{equation}
\text{Max}(\mathcal{K}):=\text{Max}\left(\sum_{i=0}P(\Pi_i=1)\right)
= \sqrt{5}, 
\label{eq:max_quant_kcbs} 
\end{equation} 
which is greater than the non-contextual bound.  This is an indication of contextual advantage of
quantum probability distributions.

It is to be noted that any valid construction of the KCBS
scenario in any formalism must necessarily ensure the
exclusivity conditions~\eqref{eq:exclusivity}.
\begin{figure}
\includegraphics[scale=1]{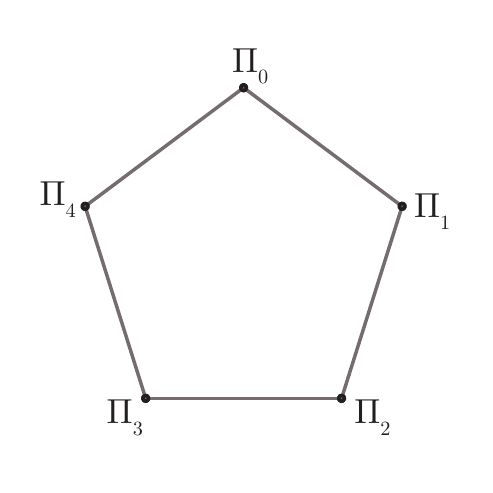}
\caption{The KCBS scenario with $5$ projectors. Projectors
connected by an edge are orthogonal. }
\end{figure}


\section{ABL rule and KCBS inequality}
\label{abl_kcbs}
The foremost requirement to check whether the
non-paradoxical sector of ABL formalism offers any
contextual advantage is to setup the KCBS scenario with the
proper exclusivity conditions given in
Eq.~\eqref{eq:exclusivity} by assigning a probability
distribution $\zeta_i$ to the projectors under a PPS
scenario. We choose the PPS as $|\psi\rangle$ and
$|\phi\rangle$ respectively to assign a counterfactual
probability distribution to the projector $\Pi_i$ according
to the ABL rule as
\begin{equation}
\zeta_i =
\frac{|\langle\phi|\Pi_i|\psi\rangle|^2}{|\langle\phi|\Pi_i|\psi\rangle|^2+
|\langle\phi|(\mathds{1}-\Pi_i)|\psi\rangle|^2},
\label{eq:ablrule}
\end{equation}
where we the measurement setting is of the form
$\lbrace\Pi_i,\mathds{1}-\Pi_i\rbrace$.

By careful
selection of a PPS, such counterfactual assignments can lead
to a logical paradox in which $\Pi_i$ and
$\Pi_{i\oplus 1}$ are assigned probabilities leading to
$\zeta_i+\zeta_{i\oplus 1}>1$.  This is a direct violation
of the exclusivity
conditions~(\ref{eq:exclusivity}). Furthermore, in order to
analyse the non-paradoxical sector of the KCBS scenario, it
is required that $\zeta_i+\zeta_{i\oplus 1}\leq 1$ for all
two-states $\bra{\phi} \ket{\psi}$.

We now propose the following setup to test the KCBS
inequality using the ABL formalism. Without loss of
generality we assume a pre-selected state
$|\psi\rangle$ as
\begin{equation}
|\psi\rangle = (0,0,1)^T,
\end{equation}
and a post-selected state as
\begin{equation}
|\phi\rangle = (\cos\theta, \sin\theta\cos\phi,\sin\theta\sin\phi)^T,
\label{eq:post_selected}
\end{equation}
where $\theta\in[0,\pi]$ and $\phi\in[0,2\pi]$. The
projectors $\Pi_i=|v_i\rangle\langle v_i|$ are of the form,
\begin{equation}
\begin{aligned}
|v_0\rangle &= \left(1,0,\sqrt{\cos\pi/5}\right)^T,\\
|v_1\rangle &= \left(\cos 4\pi/5,-\sin
4\pi/5,\sqrt{\cos\pi/5}\right)^T,\\
|v_2\rangle & = \left(\cos 2\pi/5,\sin
2\pi/5,\sqrt{\cos\pi/5}\right)^T,\\
|v_3\rangle &= \left(\cos 2\pi/5,-\sin
2\pi/5,\sqrt{\cos\pi/5}\right)^T,\\
|v_4\rangle & = \left(\cos 4\pi/5,\sin
4\pi/5,\sqrt{\cos\pi/5}\right)^T.
\end{aligned}
\label{eq:projectors}
\end{equation}

We then optimize $|\phi\rangle$ for maximum value of
$\mathcal{K}$ with the exclusivity
conditions~(\ref{eq:exclusivity}) imposed on $\zeta_i$. We
evaluate $\mathcal{K}$ using the rule~(\ref{eq:ablrule}) for
the measurement $\lbrace\Pi_i,\mathds{1}-\Pi_i\rbrace$ to
assign $P(\Pi_i =1)=\zeta_i$ accordingly. 

\begin{figure} \includegraphics[scale=1]{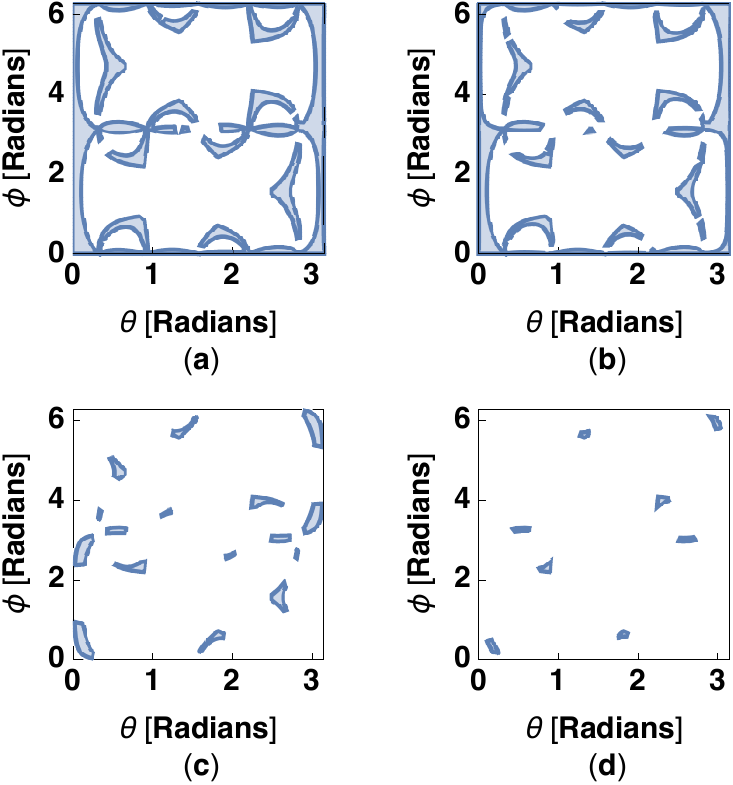}
\caption{A region plot corresponding to a set of
post-selected states~(\ref{eq:post_selected}) (shaded blue)
for which Eq.~\eqref{eq:exclusivity} is satisfied $\forall
i$ with (a) $\mathcal{K}> 1.4$, (b) $\mathcal{K}>1.5$, (c)
$\mathcal{K}>1.6$ and (d) $\mathcal{K}>1.7$. No set of
states were found for $\mathcal{K}>2.0$.}
\label{fig:intersection} \end{figure}

By imposing the exclusivity conditions we found that no post-selection can
lead to a violation of the KCBS inequality.  In
Fig~\ref{fig:intersection} we plot the intersection of the
solutions that satisfy the
inequalities~\eqref{eq:exclusivity} and \eqref{eq:kcbs} for
various minimum values of $\mathcal{K}$ over the entire
region of post-selected states. Incidentally we do not find
any set of states for which KCBS inequality is violated.
This provides a clear evidence of the fact that the ABL
formalism does not provide a complete description under the
non-paradoxical sector of PPS scenario. 

In order to achieve
a violation it is necessary to violate at least a single
exclusivity constraint. If all the constraints are
satisfied, the resultant distribution, even from the ABL
retrodiction formula is necessarily non-contextual. 

Therefore, any violation of the KCBS inequality observed via the ABL
rule, must necessarily arise from the paradoxical sector of
PPS scenarios. As a consequence, the maximum violation can
even go above  the algebraic bound. This is because, the
exclusivity conditions are not properly satisfied.

It is a natural consequence of this work that a valid
PPS-KCBS scenario can be modelled using a non-contextual
ontological model.
 
Our analysis can be extended to arbitrary contextuality
scenarios too~\cite{PhysRevA.88.022118}.  Following our
analysis, it is required to firstly identify the proper
exclusivity conditions according to
Eq.~\eqref{eq:exclusivity}. These conditions demarcate the
set of non-paradoxical PPS scenarios from the paradoxical
ones according to definition $3$. Within this set of
counterfactual PPS scenario one can then vary the
pre-selected and post-selected states for a given set of
PVMs (which define the corresponding contextuality scenario)
to make counterfactual probability assignments. A violation
of the corresponding contextuality inequality would then
indicate a contextual advantage of the TSVF. 

The KCBS scenario requires a set of $5$ PVMs, and imposes
$5$ exclusivity constraints on the ABL rule. Any $n$-cycle
scenario~\cite{PhysRevA.88.022118} would 
then consequently impose  $n$ such exclusivity constraints.
This in turn reduces the non-paradoxical 
sector of PPS scenarios possible for this contextual
inequality. While a solution for all $n$-cycle scenarios with $n$
exclusivity constraints applied to TSVF is not possible, we conjecture with good confidence that no $n$-cycle contextual
inequality can exhibit a violation under the 
TSVF paradigm.

\section{Conclusion}
\label{sec:conc}

In this work we have focused on unearthing quantum
contextuality as identified by the violation of KCBS inequality
in the PPS scenarios where ABL rule provides a way to assign 
counterfactual probabilities to measurement outcomes.
We provide a classification of PPS scenarios into paradoxical and
non-paradoxical sectors.
We then show that the
non-paradoxical sector of the ABL rule to evaluate
probability distribution over the outcomes of an observable
in a PPS scenario does not provide a contextual
violation of the KCBS inequality.
Since,  the ABL rule is applied in
a counterfactual manner, the ABL rule
acts as an
ontic model of the KCBS inequality. By imposing proper exclusivity conditions on the ABL
probabilities, we
find that it is not able to reproduce the statistics that
are observed in nature. 

It should be noted that the KCBS scenario, and the pentagram graph in general, underly 
many other contextual scenarios as well~\cite{BBC11}. Apart from KCBS, our result also 
implies a non-contextual behavior of these scenarios under the paradigm of TSVF.

Our results show that the ABL rule is essentially
non-contextual contrary to recent
studies~\cite{PhysRevLett.95.200405, PhysRevLett.113.200401,
PhysRevA.100.042116}. Most of the recent studies deal with
probability assignments which are not properly conditioned
and lead to scenarios where the sum of probabilities of
exclusive events can be greater than $1$ leading to false
signatures of contextuality. Any such signature arises from definition $2$
and therefore violates the principle of exclusivity which is at the heart of
operational theories~\cite{PhysRevLett.110.060402, PhysRevA.89.030101, PhysRevResearch.2.042001}.
Therefore, in order to observe a violation, the principle of exclusivity needs to be abandoned.

\begin{acknowledgements}
JS acknowledges support by \href{http://dx.doi.org/10.13039/100009042}{Universidad de Sevilla} Project Qdisc (Project No.\ US-15097), with FEDER funds, \href{http://dx.doi.org/10.13039/501100011033:}{MCINN/AEI} Projet No.\ PID2020-113738GB-I00, and QuantERA grant SECRET, by \href{http://dx.doi.org/10.13039/501100011033:}{MCINN/AEI} (Project No.\ PCI2019-111885-2). A.
acknowledges financial support from
DST/ICPS/QuST/Theme-1/2019/Q-68.

\end{acknowledgements}

%

\end{document}